# Real-time visualization of plasmonic nanoparticle growth dynamics by high-speed atomic force microscopy

Fuma Wakabayashi[1], Kenta Tamaki[2], Feng-Yueh Chan[2], Takayuki Uchihashi[2,3], Prabhat Verma[1], Takayuki Umakoshi[1]*

1. Department of Applied Physics, The University of Osaka, 2-1 Yamadaoka, Suita, Osaka 565-0871 Japan
2. Department of Physics, Nagoya University, Furo-cho, Chikusa-ku, Nagoya, Aichi 464-8602, Japan
3. Exploratory Research Center on Life and Living Systems, National Institutes of Natural Sciences, 5-1 Higashiyama, Myodaiji, Okazaki, Aichi 444-8787, Japan

*E-mail: umakoshi@ap.eng.osaka-u.ac.jp

**ABSTRACT**

Plasmonic nanoparticles generate strongly localized and enhanced light field through localized surface plasmon resonance, thereby playing a central role in plasmonics and nanophotonics. Because the optical properties of plasmonic nanoparticles are highly sensitive to their size and shape, nanoscale visualization of nanoparticle growth is crucial for detailed understanding of growth mechanisms and precise control of particle geometry. However, it is not possible to visualize the rapid growth dynamics using conventional imaging techniques. In this study, we demonstrate in-situ real-time observation of silver nanoparticle (AgNP) growth dynamics at the single-particle level using high-speed atomic force microscopy (HS-AFM). We employed a photoreduction method, which enables reliable control of AgNP formation by laser irradiation. By integrating a stand-alone tip-scan HS-AFM with an optical setup for photoreduction, we successfully captured real-time movies showing the nucleation and subsequent growth of AgNPs at the single-particle level. Furthermore, quantitative single-particle analysis revealed particle-to-particle variations in growth dynamics. The growth dynamics were further studied at different laser intensities, revealing intensity-dependent growth rates and the balance between nucleation and growth. This study establishes HS-AFM as a novel microscopic platform for in-situ visualization of plasmonic nanoparticle growth and will contribute to advances in plasmonics and materials science.

**Introduction**

Plasmonic nanoparticles, such as gold and silver nanoparticles, generate strongly localized and enhanced light field through localized surface plasmon resonance, and have therefore been widely employed in various plasmonic applications.[1–10] In general, plasmonic nanoparticles are chemically synthesized via the reduction of metal ions using reducing agents.[11–14] In such chemical synthesis, precise control of the size and shape of nanoparticles is crucial, as these factors strongly influence plasmonic properties such as the resonance wavelength and field enhancement factor.[1,10–12,15–18] Visualization of the growth process of plasmonic nanoparticles at the single-particle level is thus essential for a detailed understanding of their growth mechanisms and for achieving better control of the geometries of plasmonic nanoparticles. However, optical microscopy does not provide sufficient spatial resolution to precisely observe the size and shape of a single nanoparticle. In contrast, atomic force microscopy (AFM) offers nanoscale spatial resolution, enabling direct observation of individual nanoparticles. Nevertheless, its imaging speed is insufficient to track the growth process of nanoparticles. To observe the growth dynamics of individual plasmonic nanoparticles, both nanoscale spatial resolution and sufficient temporal resolution to follow the growth process are required.

In this study, we achieved in-situ real-time observation of growth dynamics of silver nanoparticles (AgNPs) at the single-particle level using high-speed AFM (HS-AFM). HS-AFM is a powerful microscopic technique that simultaneously offers high spatial and temporal resolutions, providing a spatial resolution comparable to that of conventional AFM (~1 nm) and a typical temporal resolution of >10 frames per second (fps).[19] Owing to its great imaging capability, HS-AFM has made significant contributions to biological research by visualizing dynamic motions of functional proteins under physiological conditions.[20–26] In recent years, HS-AFM has also been applied to other research fields, including materials science and organic chemistry.[27–38] In particular, HS-AFM visualization of the nucleation and growth dynamics of supramolecular assemblies has been recently reported.[34] Therefore, plasmonic nanoparticles represent a suitable system for which HS-AFM can provide unique insights into growth dynamics. To effectively observe the formation and growth of AgNPs, we employed photoreduction of silver ions.[39–42] In the photoreduction reaction, silver ions are reduced and grow into nanoparticles upon light irradiation. As we can easily switch on and off the reaction by light, the photoreduction method enables reliable control over nanoparticle growth and facilitates HS-AFM measurements. To effectively combine photoreduction with HS-AFM observation, we employed a tip-scan HS-AFM.[43,44] Unlike conventional HS-AFM, in which imaging is performed by scanning the sample stage, tip-scan HS-AFM rapidly scans the tip. Importantly, tip-scan HS-AFM is implemented as a stand-alone system so that it can be mounted on an inverted optical microscope, which greatly facilitates reliable integration with optical techniques.[32,45,46] Furthermore, evanescent light was introduced at the substrate surface via total internal reflection to induce the photoreduction reaction. Because the evanescent field is localized within a few hundred nanometers from the substrate surface, the formation and growth of AgNPs can be selectively induced only near the surface, resulting in stable and reliable HS-AFM operation. Using this developed system, we successfully captured real-time movies of the nucleation and subsequent growth of AgNPs at the single-nanoparticle level. Direct visualization of the growth process at the single-particle level provides new insights into plasmonic nanoparticle synthesis and represents a significant advance in the field of plasmonics.

**Results & discussion**

Figure 1 shows a schematic of the experimental setup for HS-AFM observation of the AgNP growth. A custom-built tip-scan HS-AFM system was installed on an inverted optical microscope. Details of the tip-scan HS-AFM system are described in previous studies.[32,43] We constructed a total internal reflection illumination setup beneath the tip-scan HS-AFM, which was similar to that used in objective-type total internal reflection fluorescence (TIRF) microscopy. A laser beam (Cobolt, Samba 0532-04-01-0100-700, $\lambda$=532 nm) was expanded using a beam expander. It was then weakly focused at the pupil plane of an oil-immersion objective lens using a lens (f = 300 mm) such that a collimated laser beam emerged from the objective lens. By introducing the focused beam through the peripheral region of the objective, the laser beam was emitted from the objective with an angle larger than the critical angle. Consequently, the laser beam was totally internally reflected and only an evanescent light field was generated on the other side of the glass substrate, which was localized near the substrate surface. This configuration confined the photoreduction reaction to the vicinity of the substrate, while suppressing photoreduction in the bulk solution away from the substrate. This is important for stable HS-AFM operation, as the formation of a large number of AgNPs in the bulk solution can disturb HS-AFM measurements. Using this configuration, we achieved stable HS-AFM observation of the nucleation and growth of AgNPs. To observe the photoreduction reaction and the formation of AgNPs, we adjusted the tip position to the center of the illumination spot. The illumination spot size was approximately 10 μm. For photoreduction of AgNPs, we prepared an aqueous solution containing eosin-Y-disodium (eosin) (10 μM), *N*-methyldiethanolamine (MDEA) (1 mM), and $AgNO_3$ (0.5 mM). Eosin acted as a photosensitizer, while MDEA served as a reducing agent to reduce silver ion supplied from $AgNO_3$. The mechanism of the photoreduction reaction is described in previous studies.[41,42] The solution was dropped onto a piranha-cleaned glass substrate, and the photoreduction reaction was observed in situ using a tip-scan HS-AFM.

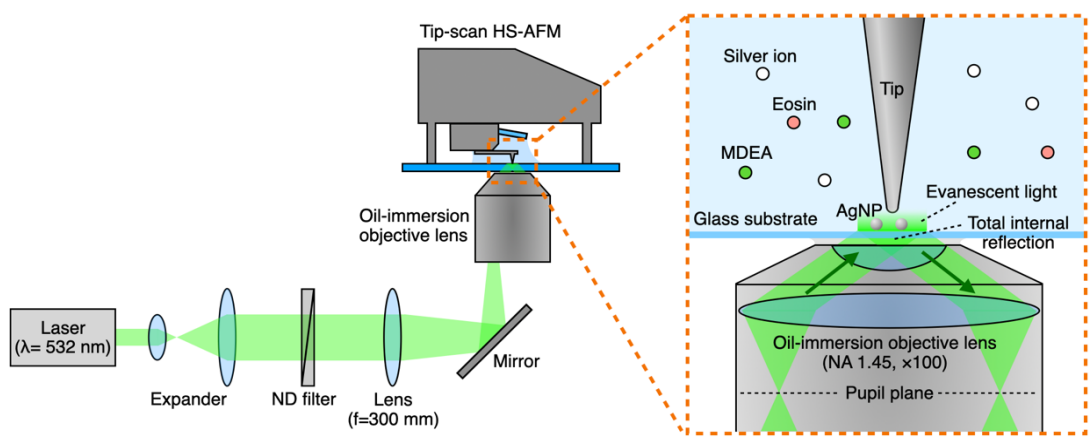

**Figure 1.** Experimental setup for HS-AFM observation of the AgNP growth.

Using the constructed experimental system, we performed in-situ HS-AFM video recording of the process of nucleation and growth of AgNPs under photoreduction at an imaging rate of 1 fps with an image size of 300 × 300 $nm^2$. Laser irradiation was initiated at 0 s with a laser intensity of 10 $W/cm^2$, and the substrate surface was continuously illuminated thereafter. As shown in Figure 2a and Supporting Movie S1, we successfully observed the nucleation of small AgNPs several seconds after the onset of laser irradiation. It should be noted that the small dark pits in the HS-AFM images are inherent features of the glass substrate and are unrelated to the AgNP formation. The number of AgNPs increased with time, indicating that nucleation proceeded continuously. In addition, the size of each AgNP increased over time due to particle growth. Figures 2b–d represent magnified HS-AFM images to clearly show individual AgNPs, corresponding to **P1**, **P2**, and **P3** indicated by the dotted squares in Figure 2a. The growth of individual AgNPs was clearly visualized over time. Because HS-AFM provides precise height information, Figure 2e presents the temporal evolution of the volumes of individual AgNPs in Figures 2b–d, estimated from the measured heights by assuming a spherical geometry and using the particle height as the diameter. The temporal evolution of the heights of the corresponding particles was also included in Supporting Figure S1a. The volumes approximately linearly increased over the observation period. The growth rates of individual AgNPs were similar, with a slight variations. Owing to the high spatiotemporal resolution of HS-AFM, the growth rate was quantitatively estimated to be approximately 1.8 nm$^3$/s. Furthermore, the number of AgNPs within the field of view of HS-AFM was counted, as shown in Figure 2f. The particle number initially increased monotonically, and then interestingly, it was gradually saturated at approximately 200 particles at around 80 s. This behavior indicates that the nucleation and growth process occurred simultaneously when the particle density was low, and the growth process became dominant beyond a certain particle density because the particle growth is energetically more favorable than further nucleation.

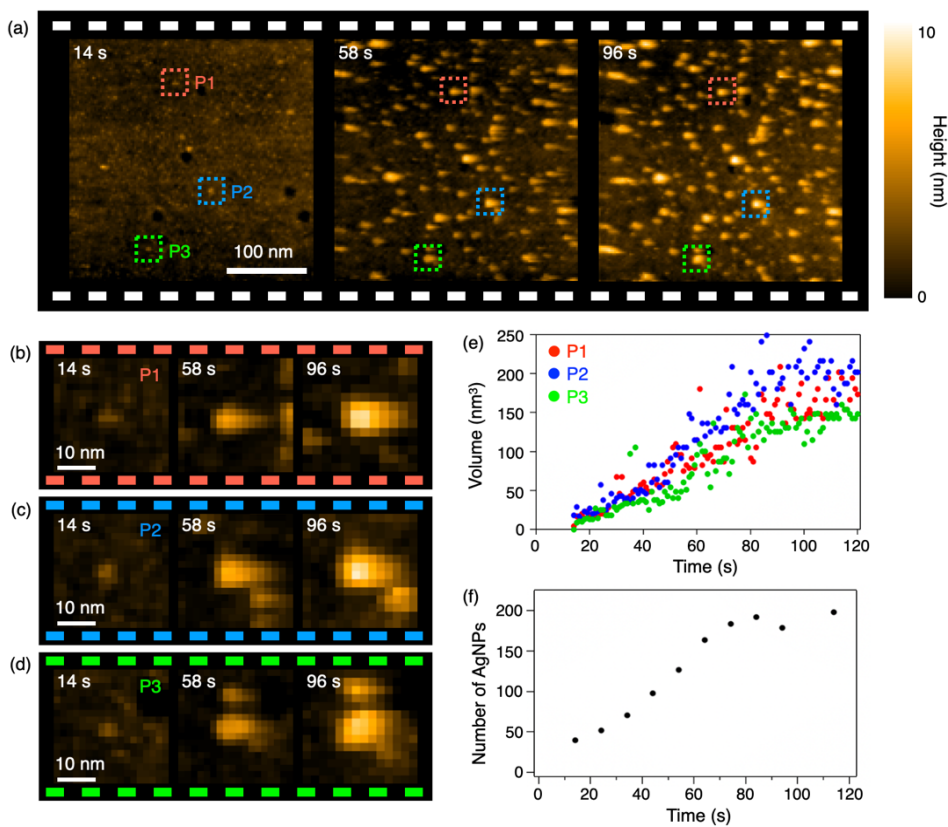

**Figure 2.** (a) Clipped HS-AFM images showing the growth of AgNPs at a laser intensity of 10 W/cm$^2$. (b-d) Magnified HS-AFM images of AgNPs **P1**, **P2**, and **P3**, respectively, corresponding to the red, blue, and green dotted squares in (a). (e) Temporal evolution of the volume of the corresponding AgNP **P1**, **P2**, and **P3**. (f) Temporal evolution of the number of AgNPs within the HS-AFM field of view.

Next, we investigated the growth process of AgNPs at a higher laser intensity (60 W/cm$^2$), as shown in Figure 3a and Supporting Movie S2. In this case, instability in HS-AFM operation and mechanical drift were observed for around 50 s immediately after the laser irradiation, probably due to the thermal drift caused by intense incident laser that perturbed the HS-AFM imaging conditions. In addition, optical radiation pressure from the incident laser may have contributed to this instability. Therefore, tracking of identical AgNPs was not possible before 50 s. Nevertheless, it was still possible to count the number of particles in the field of view before 50 s, and we could still observe that nucleation proceeded rapidly immediately after laser irradiation. After stabilization of the HS-AFM imaging, we further observed a gradual increase in the sizes of individual AgNPs, which is also clearly shown in the magnified HS-AFM images in Figures 3b–d. Single-particle analysis of the volume evolution is shown in Figure 3e, where the volume was estimated from the height evolution shown in Supporting Figure S1b. It should be noted that this single-particle analysis was performed using data acquired after 50 s because of the initial mechanical drift mentioned above. The AgNPs exhibited a growth behavior similar to that observed at a laser intensity of 10 W/cm$^2$ (Figure 2e). However, individual AgNPs showed slightly different growth dynamics. The volume of AgNPs **P1** and **P2** increased with growth rates of approximately 2.2 nm$^3$/s, whereas AgNP **P3** exhibited a faster growth rate of 3.8 nm$^3$/s. Interestingly, at around 170 s, the growth rates of AgNP **P1** and **P3** increased to 8.3 and 7.3 nm$^3$/s, respectively, while the growth rate of AgNP **P2** remained constant during the observation period. Although the detailed mechanism underlying these differences is still under investigation, this behavior can be attributed to variations in the local environment surrounding each AgNP. For example, differences in the spatial distribution of neighboring AgNPs, the internal molecular packing order of each AgNP, and the local concentration of silver ions around individual particles can influence growth dynamics. In addition, nearby small silver nuclei may have merged to form a larger AgNP, thereby increasing the apparent growth rate. Owing to the powerful imaging capability of HS-AFM, such heterogeneous growth behavior of individual AgNPs was quantitatively visualized at the nanoscale. We also counted the number of AgNP within the HS-AFM field of view (Figure 3f). Although the particle number first increased, it was almost saturated at 60 s, which was earlier than that observed at 10 W/cm$^2$. Moreover, the number of particle was eventually approximately 100, which was much fewer than the case of laser intensity of 10 W/cm$^2$. We attribute this behavior to faster particle growth at higher laser intensity, which reduces the local concentration of silver ions available for new nucleation events, thereby decreasing the probability of nucleation and resulting in a lower particle density.

Finally, we further increased the laser intensity to 480 W/cm$^2$. Under this condition, nucleation and growth of AgNPs occurred extremely rapidly, and the substrate surface was covered with a large number of AgNPs within a few seconds, as shown in Figure 4 and Supporting Movie S3. Therefore, statistical and quantitative analysis of the growth processes of individual particles was not feasible. Nevertheless, we observed a remarkably high growth rate and a very high density of AgNPs occupying the substrate surface at this high laser intensity.

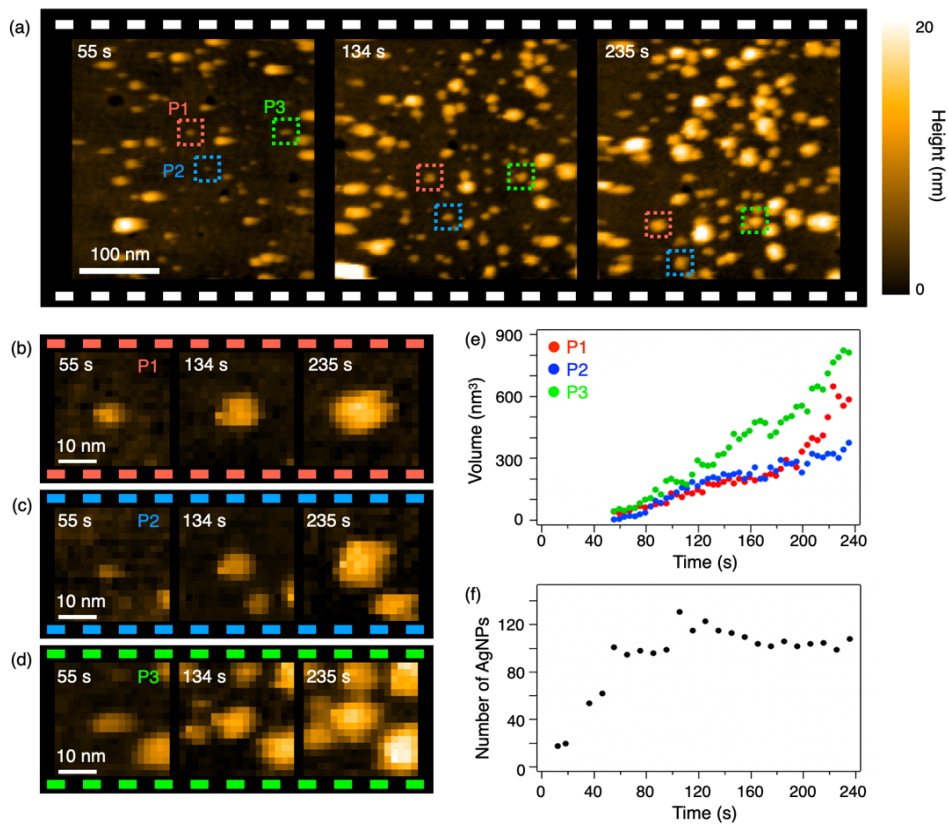

**Figure 3.** (a) Clipped HS-AFM images showing the growth of AgNPs at a laser intensity of 60 W/cm$^2$. (b-d) Magnified HS-AFM images of AgNPs **P1**, **P2**, and **P3**, respectively, corresponding to the red, blue, and green dotted squares in (a). (e) Temporal evolution of the volume of the corresponding AgNP **P1**, **P2**, and **P3**. (f) Temporal evolution of the number of AgNPs within the HS-AFM field of view.

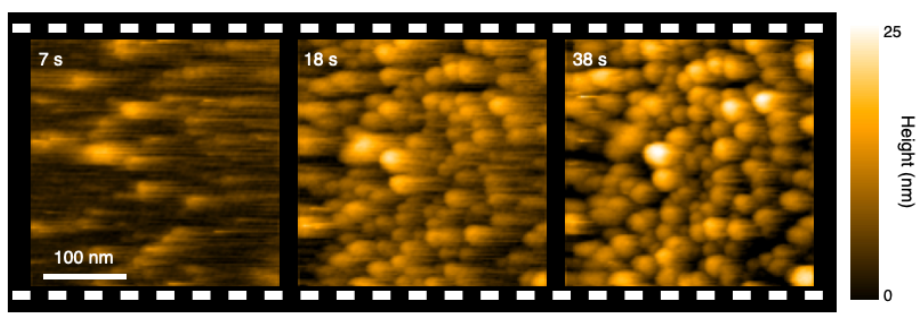

**Figure 4.** Clipped HS-AFM images showing the growth of AgNPs at a laser intensity of 480 W/cm$^2$.

## Conclusion

In this study, we achieved in-situ real-time visualization of the nucleation and growth of AgNPs at the single-particle level using HS-AFM. By integrating tip-scan HS-AFM with evanescent-field-induced photoreduction, we captured real-time movies of nanoparticle formation and subsequent growth with high spatiotemporal resolution. Quantitative analyses of individual particles revealed particle-to-particle variations in growth dynamics, highlighting the significant imaging capability of HS-AFM to directly probe growth processes that are difficult to access using conventional characterization methods. At present, HS-AFM measurements were performed under limited reaction conditions. Further studies under a variety of reaction conditions will contribute to elucidating the detailed growth mechanism of plasmonic nanoparticles. For example, we can extensively study the dependence of growth dynamics on the concentrations of silver ions, photosensitizer, and reducing agents. Moreover, by adding capping agents for shape control, observation of plasmonic nanostructures with different shapes, such as silver nanorods, would provide valuable insights. Extending this approach to other plasmonic materials, such as gold and aluminum, will also deepen our understanding of plasmonic nanoparticle growth. Overall, HS-AFM enables direct nanoscale visualization of plasmonic nanoparticle growth, establishing a promising microscopic platform for elucidating growth mechanisms and advancing plasmonics and materials science.

**Conflicts of interest**

The authors have no conflicts of interest to declare.

**Acknowledgements**

This work was partially supported by JST FOREST (JPMJFR233Z), JSPS Grant-in-Aid for Scientific Research (KAKENHI) (Grant in Aid for Scientific Research (B) JP24K01385, Grant-in-Aid for Transformative Research Areas (A) Publicly Offered Research "Chiral materials science pioneered by the helicity of light" JP25H01624, Grant-in-Aid for Transformative Research Areas (A) Publicly Offered Research "Materials science of meso-hierarchy" JP24H01717, Grant-in-Aid for Challenging Research (Exploratory) JP24K21718), and a research grant from the Takahashi Industrial and Economic Research Foundation.

*Supporting Information*

# Real-time visualization of plasmonic nanoparticle growth dynamics by high-speed atomic force microscopy

Fuma Wakabayashi[1], Kenta Tamaki[2], Feng-Yueh Chan[2], Takayuki Uchihashi[2,3], Prabhat Verma[1], Takayuki Umakoshi[1]*

1. Department of Applied Physics, The University of Osaka, 2-1 Yamadaoka, Suita, Osaka 565-0871 Japan
2. Department of Physics, Nagoya University, Furo-cho, Chikusa-ku, Nagoya, Aichi 464-8602, Japan
3. Exploratory Research Center on Life and Living Systems, National Institutes of Natural Sciences, 5-1 Higashiyama, Myodaiji, Okazaki, Aichi 444-8787, Japan

*E-mail: umakoshi@ap.eng.osaka-u.ac.jp

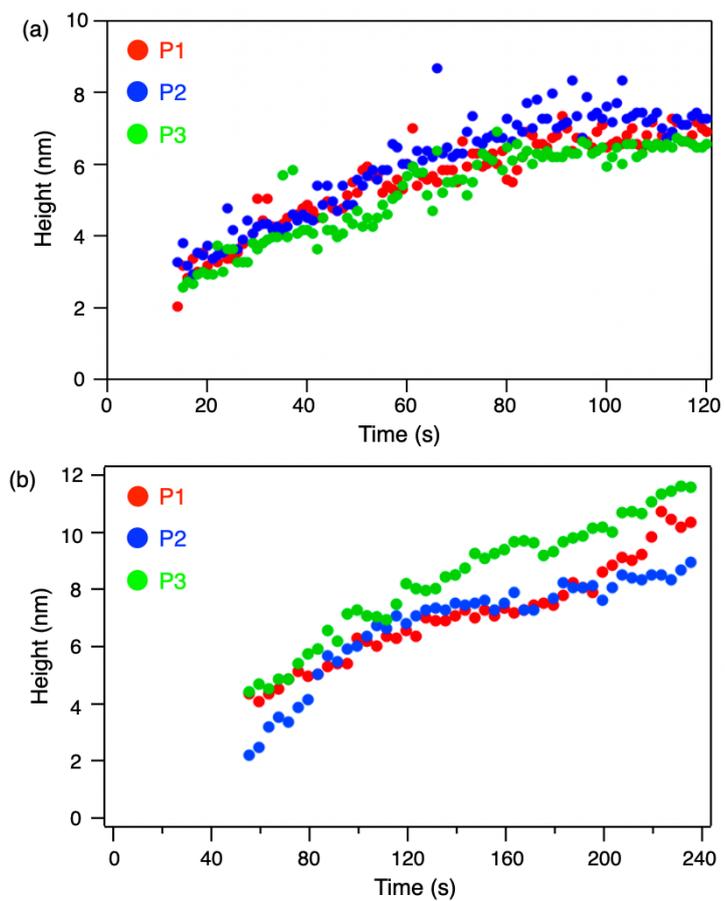

**Supporting Figure S1.** (a) Temporal evolution of the heights of AgNP **P1**, **P2**, and **P3** in Fig. 2(a). (b) Temporal evolution of the heights of AgNP **P1**, **P2**, and **P3** in Fig. 3(a).

**Supporting Movie S1:** https://youtube.com/shorts/2z3V1HSYu6M

HS-AFM movie showing the nucleation and growth dynamics of AgNPs formed via photoreduction at a laser intensity of 10W/cm$^2$, captured in a 300 × 300 nm$^2$ range, 128 × 128 pixels, at 1 fps.

**Supporting Movie S2:** https://youtube.com/shorts/8-YsFM0Om34

HS-AFM movie showing the nucleation and growth dynamics of AgNPs formed via photoreduction at a laser intensity of 60W/cm$^2$, captured in a 300 × 300 nm$^2$ range, 128 × 128 pixels, at 1 fps.

**Supporting Movie S3:** https://youtube.com/shorts/-0pB9IrycJA

HS-AFM movie showing the nucleation and growth dynamics of AgNPs formed via photoreduction at a laser intensity of 480W/cm$^2$, captured in a 300 × 300 nm$^2$ range, 128 × 128 pixels, at 1 fps.